\documentclass[aps,prl,twocolumn,showpacs]{revtex4}
\usepackage{graphicx}
\usepackage{epsf}
\usepackage{psfrag}
\usepackage[T1]{fontenc} 
\usepackage{epsfig,latexsym}
\usepackage{color}
\usepackage[latin1]{inputenc}
\usepackage{amsmath}
\usepackage{amssymb}
\usepackage{amsfonts}
\usepackage{indentfirst}
\usepackage{ae}
\usepackage{float}

\newcommand{\bq}{\begin{eqnarray}}
\newcommand{\eq}{\end{eqnarray}}

\begin{document}
\title{Gouy phase in non-classical paths in triple-slit interference experiment}
\author{I. G. da Paz$^{1}$, C. H. S. Vieira$^{1}$, R. Ducharme$^{2}$, L. A. Cabral$^{3}$, H. Alexander$^{4}$, M. D. R. Sampaio$^{4}$}

\affiliation{$^1$ Departamento de F\'{\i}sica, Universidade Federal
do Piau\'{\i}, Campus Ministro Petr\^{o}nio Portela, CEP 64049-550,
Teresina, PI, Brazil}

\affiliation{$^{2}$ 2112 Oakmeadow Pl., Bedford, TX 76021, USA}

\affiliation{$^{3}$ Curso de F\'{\i}sica, Universidade Federal do
Tocantins, Caixa Postal 132, CEP 77804-970, Aragua\'{\i}na, TO,
Brazil}

\affiliation{$^{4}$ Departamento de F\'{\i}sica, Instituto de
Ci\^{e}ncias Exatas, Universidade Federal de Minas Gerais, Caixa
Postal 702, CEP 30161-970, Belo Horizonte, Minas Gerais, Brazil}

\begin{abstract}
We propose a simple model to study the Gouy phase effect in the
triple-slit experiment in which we consider a non-classical path.
The Gouy phase differs for classical or non-classical paths as it
depends on the propagation time. In this case the Gouy phase
difference changes the Sorkin parameter $\kappa$  used to estimate
non-classical path contribution in a nontrivial way shedding some
light in the implementation of experiments to detect non-classical
path contributions.
\end{abstract}

\pacs{41.85.-p, 03.65.Ta, 42.50.Tx, 31.15.xk}

\maketitle

\section{I. Introduction}
The Gouy phase shift in light optics was theoretically studied and
experimentally observed by L. G. Gouy in 1890 \cite{gouy1,gouy2}.
The physical origin of this phase was studied  in
\cite{Visser,simon1993,feng2001,yang,boyd,hariharan,feng98, Pang}.
The Gouy phase shift appears in any kind of wave that is submitted
to transverse spatial confinement, either by focusing or by
diffraction through small apertures. When a wave is focused
\cite{feng2001}, the Gouy phase shift is associated to the
propagation from $-\infty$ to $+\infty$ and is equal to $\pi/2$ for
cylindrical waves (line focus), and $\pi$ for spherical waves (point
focus). In the case of diffraction by a slit it was shown that the
Gouy phase shift is $\pi/4$ and it is dependent on the slit width
and the propagation times before and after the slit \cite{Paz2}. The
Gouy phase shift has been observed in different kind of waves such
as water waves \cite{chauvat}, acoustic \cite{holme}, surface
plasmon-polariton \cite{zhu}, phonon-polariton \cite{feurer} pulses,
and recently in matter waves \cite{cond,elec2,elec1}.

Applications of Gouy phase in light optics opening the possibility
of development of optical systems has been the subject of many studies
and increasing interest. For instance, the Gouy phase has to be taken
into account to determine the resonant frequencies in laser cavities
\cite{siegman} or the phase matching in high-order harmonic generation
(HHG) \cite{Balcou} and to describe  the spatial variation of the carrier
envelope phase of ultrashort pulses in a laser focus \cite{Lindner}.
Moreover, the Gouy phase plays important role in the evolution of
optical vortex beams \cite{Allen} as well as electron beams which
acquire an additional Gouy phase dependent on the absolute value of
the orbital angular momentum \cite{elec2}. Gravity wave
detection antennas are based on precision measurements using laser
interferometry in which the Gouy phase is crucial \cite{Sato}.

In the non-relativistic matter wave context the Gouy phase has been
explored firstly in \cite{PNP,Paz1}, followed by experimental
realizations with Bose-Einstein condensates \cite{cond}, electron
vortex beams \cite{elec2} and astigmatic electron matter waves using
in-line holography \cite{elec1}. Recently, it was showed that the
Beteman-Hillion solutions to the Klein-Gordon equation presents a
Gouy phase that includes relativistic effects \cite{Ducharme}. Matter
wave Gouy phases have interesting applications as well. For instance, they serve as
mode conversers important in quantum information \cite{Paz1},
in the development of singular electron optics \cite{elec1}, in studying the
Zitterbewegung phenomenon \cite{Ducharme}, and now we investigate
how important it can be in the study of non-classical paths in
interference experiments such as less likely, more exotic, looping paths  as we shall
explain below. From the theoretical viewpoint,  the contribution of such exotic trajectories
amounts to saying that the superposition principle is usually incorrectly applied in interference
experiments.

A theoretical treatment of non-classical path in the double-slit was
studied in \cite{Yabuki}. They estimated  a nonlinear interference
term to test a deviation from the superposition principle in the
double slit experiment. They used the Feynman path integral approach
\cite{FeynmanHibbs} with inclusion of paths looping along the slits,
i.e., non-classical paths. Experimental access to such tiny
deviations was discussed  by  Sorkin \cite{Sorkin} in a work where
higher-order phenomena incorporate the usual prescription of
interference when three or more paths interfere. However, only
recently was proposed a quantification of the non-classical paths in
interference experiments for triple-slit
\cite{Sinha1,Raedt,Sinha2,Sinha3}. The theoretical analysis to
support these experiments are based in path integrals in the
presence of slits with different weights for classical and
non-classical paths, namely the propagator is written as
$$K(\vec{r}_1,\vec{r}_2)=\int {\cal{D}} [\vec{x}(s)] \exp[i k \int
ds],$$ where $s$ is the contour length along $\vec{x}(s)$, the
classical limit being $k\rightarrow \infty$ where paths near the
straight line linking $\vec{r}_1$ to $\vec{r}_2$ contribute by
stationary phase. Paths away from the classical path contribute with
a rapidly oscillating phase. All paths from source to detector
should be considered excluding those who would cross the opaque
walls along the slits.

In \cite{Sinha1} it was introduced the Sorkin factor $\kappa$ which
gauges the deviation of the Born rule for probabilities in quantum
mechanics, i.e. to estimate contributions from non-classical paths.
$\kappa = 0$ if only classical paths contribute to final
interference pattern in detector and $\kappa \neq 0$ if, beyond
usual classical paths, non-classical paths are considered in the
calculations and contribute to final result. For the usual
double-slit experiment, until the present time it was not detected
any deviation from a null value of $\kappa$. However new experiments
with three slits proposed in \cite{Sinha3} using matter waves or low
frequency photons were analytically described enabling to set an
upper bound on the Sorkin factor $|\kappa_{max}| \approx 0.003
\lambda^{3/2}/(d^{1/2} w)$, in which $\lambda$ is the wavelength,
$d$ is the center to center distance between the slits and $w$ is
the width of the slit. They  confirmed that $\kappa$ is very
sensitive to the experimental setup.

The guiding purpose of this manuscript is to incorporate the effect
of the Gouy phase into parameter $\kappa$ and indicate this effect
on the pattern of interference as well in $\kappa$ for matter waves.
As we shall see, the Sorkin factor for triple-slit interference is
dependent on the Gouy phase difference between classical and
non-classical paths. The effect of Gouy phase of matter waves has
recently earned prominence with its inclusion in electron beams
which are used in  \cite{Sinha2}, \cite{Sinha3} to estimate $\kappa
\approx 10^{-8}$. In order to analytically evaluate the interference
pattern we establish a procedure similar to that presented in
\cite{Paz2,solano} using non-relativistic propagators for a free
particle Gaussian wavepacket adapted to triple-slit interference
with non-classical paths. This framework allows for exact
integration and analytical expressions which depend on the geometry
of the experimental setup and source parameters. Moreover, we  make
explicit the Gouy phase in the wavefunctions for a triple-slit
apparatus $\psi_{1}$, $\psi_{2}$, $\psi_{3}$ and $\psi_{nc}$
(corresponding wavefunction for non-classical path) and derive an
expression for $\kappa$ which is of order $10^{-8}$ for electron
waves.

This contribution is organized as follows: in section II we obtain
analytical expressions for the wavefunctions for classical and
non-classical paths and calculate the intensity. We estimate the
deviations produced by non-classical path through the Sorkin
parameter $\kappa$. In section III we analyse the effect of the Gouy
phase in the Sorkin parameter for electron waves and we estimate the
percentage error in this parameter when we neglect the Gouy phase
difference of classical and non-classical paths in order to get some
insight in the relative importance of such effects in the
interference pattern. We draw some concluding remarks on section IV.

\section{II. Non-classical paths in triple-slit experiment}

In this section we obtain analytical results for the wave functions
of classical and non-classical paths in the triple-slit experiment
keeping track of  phases in order to assess their role in the
interference pattern. Suppose an one dimensional model in which
quantum effects are manifested only in the $x$-direction as depicted
in Fig. 1. A coherent Gaussian wavepacket of initial transverse
width $\sigma_{0}$ is produced in the source $S$ and propagates
during a time $t$ before arriving at a triple-slit with gaussian
apertures from which Gaussian wavepackets propagate. After crossing
the grid the wavepackets propagate during a time $\tau$ before
arriving at detector $D$ in detection screen giving rise to a
interference pattern as a function of the transverse coordinate $x$.
The summation over all possible trajectories allows for exotic paths
such as the one depicted in Fig. 1. We calculate the corresponding
wavefunction for this path in order to analyse its effect in the
interference pattern.

\begin{figure}[htp]
\centering
\includegraphics[width=8.0 cm]{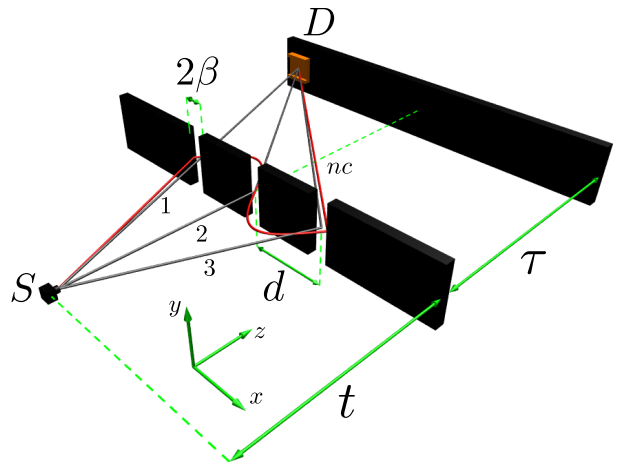}
\caption{Sketch of triple-slit experiment. Gaussian wavepacket of
transverse width $\sigma_{0}$ produced in the source $S$ propagates
a time $t$ before attaining the triple-slit and a time $\tau$ from
the triple-slit to the detector $D$. The slit apertures are taken to
be Gaussian of width $\beta$ separated by a distance
$d$.}\label{Figure1}
\end{figure}

The wave functions corresponding to the classical paths (grey lines)
$1$ and $3$ read ($\int_{-\infty}^{+\infty} \ldots
\int_{-\infty}^{+\infty} dx_1 \ldots dx_n \equiv \int_{x_1 \ldots
x_n}$):

\bq \psi_{1,3}(x,t,\tau) &=&\int_{x_j,x_0}
K_{\tau}(x,t+\tau;x_{j},t)F(x_{j}\pm
d)\nonumber \\
&\times&K_{t}(x_{j},t;x_{0},0)\psi_{0}(x_{0}), \label{psi13}
\eq
whereas for the classical path $2$
\bq
\psi_{2}(x,t,\tau) &=&\int_{x_j,x_0} K_{\tau}(x,t+\tau;x_{j},t)F(x_{j}) \nonumber \\ &\times&
K_{t}(x_{j},t;x_{0},0)\psi_{0}(x_{0}),
\label{psi2}
\eq
with
\begin{equation}
K(x_{j},t_{j};x_{0},t_0)=\sqrt{\frac{m}{2\pi i\hbar
(t_{j}-t_{0})}}\exp\left[\frac{im(x_{j}-x_{0})^{2}}{2\hbar
(t_{j}-t_0)}\right],
\end{equation}
\begin{equation}
F(x_{j})=\exp\left[-\frac{(x_{j})^{2}}{2\beta^{2}}\right],
\end{equation}
and
\begin{equation}
\psi_{0}(x_{0})=\frac{1}{\sqrt{\sigma_{0}\sqrt{\pi}}}\exp\left(-\frac{x_{0}^{2}}{2\sigma_{0}^{2}}\right).
\end{equation}
The kernels $K_{t}(x_{j},t;x_{0},0)$ and
$K_{\tau}(x,t+\tau;x_{j},t)$ are the free propagators for the
particle, the functions $F(x_{j})$ describe the slit transmission
functions which are taken to be Gaussian of width $\beta$ separated
by a distance $d$; $\sigma_{0}$ is the effective width of the
wavepacket emitted from the source $S$, $m$ is the mass of the
particle, $t$ ($\tau$) is the time of flight from the source
(triple-slit) to the triple-slit (screen). The wavefunction
associated with the non-classical path (red line) is given by
\begin{eqnarray}
\psi_{nc}(x,t,\tau)&=&\int_{x_0,x_1,x_2,x_3}
K_{\tau}(x,\tau+\tilde{t};x_{3},\tilde{t})\nonumber\\
&\times&F(x_{3}+ d)F(x_{2})K(1\rightarrow2;2\rightarrow3)\nonumber\\
&\times& F(x_{1}- d)K_{t}(x_{1},t+\alpha;x_{0},0)\psi_{0}(x_{0}),
\label{psiNC}
\end{eqnarray}
where $\tilde{t} = t+2(\epsilon +\alpha)$ and

\bq K(1\rightarrow2;2\rightarrow3)=\sqrt{\frac{m}{4\pi
i\hbar(\epsilon+\alpha)}} \times \nonumber \\
\exp\left[\frac{im[(x_{2}-x_{1})^{2}+(x_{3}-x_{2})^{2}]}{4\hbar(\epsilon+\alpha)}\right],
\eq is the free propagator which propagates from slit $1$ to slit
$2$ and from slit $2$ to slit $3$. The parameter $\alpha \rightarrow
0$ is an auxiliary inter slit time parameter and $\epsilon$ is the
time spent from one to the next slit and is determined by the
momentum uncertainty in the $x$-direction, i.e.,
$\epsilon=\frac{d}{\Delta v_{x}}$ ($\Delta v_{x}=\Delta p_{x}/m$),
where $\Delta
p_{x}=\sqrt{\langle\hat{p}^{2}_{x}\rangle-\langle\hat{p}_{x}\rangle^{2}}$,
$\hat{p}_{x}$ being the momentum operator in the $x$-direction. This
estimation is compatible with the propagation which builds the
non-classical trajectory. A similar argument was used  in
\cite{Vedral}, where non-classical dynamics based on uncertainty
principle are considered in a interferometer. Trajectories winding
around the slits evidently contribute less and less to the
interference pattern.

After some lengthy algebraic manipulations, we obtain
\begin{eqnarray}
\psi_{1}(x,t,\tau)=A\exp(-C_{1}x^{2}-C_{2}x+C_{3}) \times \nonumber \\
\exp(i\alpha x^{2}-i\gamma x-i\theta_{c} + i\mu_{c}), \label{psi1}
\end{eqnarray}

\begin{equation}
\psi_{2}(x,t,\tau)=A\exp(-C_{1}x^{2})\exp(i\alpha x^{2} + i\mu_{c}),
\label{psi2}
\end{equation}

\begin{eqnarray}
\psi_{3}(x,t,\tau)=A\exp(-C_{1}x^{2}+C_{2}x+C_{3}) \times \nonumber \\
\exp(i\alpha x^{2}+i\gamma x-i\theta_{c} + i\mu_{c}), \label{psi3}
\end{eqnarray}
and
\begin{eqnarray}
\psi_{nc}(x,t,\tau)=A_{nc}\exp(-C_{1nc}x^{2}+C_{2nc}x+C_{3nc}) \times \nonumber \\
\exp\left(i\alpha_{nc}
x^{2}+i\gamma_{nc}x-i\theta_{nc}+i\mu_{nc}\right) \label{psi_nc},
\end{eqnarray}
where the non trivial Gouy phase  $\mu_{nc}$ is given by
\begin{equation}
\mu_{nc}(t,\tau)=\frac{1}{2}\arctan\left(\frac{z_{I}}{z_{R}}\right).
\label{ncgouy}
\end{equation}
All the coefficients present in equations
(\ref{psi1})-(\ref{ncgouy}) are written out in appendices 1 and 2
for the sake of clarity. The indices $R$ and $I$ stand for the real
and imaginary part of the complexes numbers that appear in the
solutions. As discussed in \cite{Paz2}, $\mu_{nc}(t,\tau)$ and
$\theta_{nc}(t,\tau)$ are phases that do not depend of the
transverse position $x$. Different from the Gouy phase,
$\theta_{nc}(t,\tau)$ is one phase that appears as we displace the
slit from a given distance of the origin, which is dependent on the
parameter $d$.

The intensity at the screen including non-classical path reads
\begin{eqnarray}
I_{nc}&=&|\psi_{1}+\psi_{2}+\psi_{3}+\psi_{nc}|^{2}\nonumber\\
&=&I_{c}+|\psi_{nc}|^{2}+2|\psi_{1}||\psi_{nc}|\cos\phi_{1nc}\nonumber\\
&+&2|\psi_{2}||\psi_{nc}|\cos\phi_{2nc}+2|\psi_{3}||\psi_{nc}|\cos\phi_{3nc},
\label{int_trifenda}
\end{eqnarray}
where
\begin{equation}
\phi_{1nc}=(\alpha-\alpha_{nc})x^{2}-(\gamma+\gamma_{nc})x-(\theta_{c}-\theta_{nc})+(\mu_{c}-\mu_{nc}),
\label{phi1}
\end{equation}
\begin{equation}
\phi_{2nc}=(\alpha-\alpha_{nc})x^{2}-\gamma_{nc}x+\theta_{nc}+(\mu_{c}-\mu_{nc}),
\label{phi2}
\end{equation}
and
\begin{equation}
\phi_{3nc}=(\alpha-\alpha_{nc})x^{2}+(\gamma-\gamma_{nc})x-(\theta_{c}-\theta_{nc})+(\mu_{c}-\mu_{nc})
\label{phi3}
\end{equation}
are the relative phases of $\psi_{1}$ and $\psi_{nc}$, $\psi_{2}$
and $\psi_{nc}$ and $\psi_{3}$ and $\psi_{nc}$, respectively, which
implies that the interference is a result of two-paths as observed
in \cite{Park}. $I_{c}$ is the intensity when only classical paths
are taken into account.

To quantify the deviations in the intensity produced by the
existence of non-classical paths we use the Sorkin parameter as
defined in Refs. \cite{Sinha2,Sinha3}, i.e.,
\begin{eqnarray}
\kappa I_{0} &=&I_{nc}-I_{c}\nonumber\\
&=&|\psi_{nc}|^{2}+2|\psi_{1}||\psi_{nc}|\cos\phi_{1nc}\nonumber\\
&+&2|\psi_{2}||\psi_{nc}|\cos\phi_{2nc}+2|\psi_{3}||\psi_{nc}|\cos\phi_{3nc},
\label{fatorK}
\end{eqnarray}
where $I_{0}$ is the intensity at central maximum. As we can observe
the parameter $\kappa$ used to estimate the existence of
non-classical path in the triple-slit interference is dependent of
the Gouy phase difference between classical and non-classical paths.

\section{III. Sorkin parameter and Gouy phase for electron waves}

In this section we analyse  the Gouy phase effect in the quantity
$\kappa$ for electron waves. Firstly we observe the behavior of the
normalized intensity and the parameter $\kappa$ as a function of $x$
fixing the values of $t$ and $\tau$. We observe a displacement in
the behavior of $\kappa$ as an effect of the Gouy phase which make
clear the role of this phase in the exact estimation of $\kappa$.
Secondly we observe the behavior of the parameter $\kappa$ as a
function of $x$ and $\tau$ fixing the value of $t$ in which we can
observe an upper bound for a given value of $x$ and $\tau$. Thirdly
we consider the position $x=0$ and observe the behavior of the
parameter $\kappa$ as a function of $\tau$ for $t$ fixed. For $x=0$
the Gouy phase effect is most evident since some other phases
disappear in the interference as we can see in equations
(\ref{phi1})-(\ref{phi3}). As a matter of fact we can study the
effect of all phases that appear with the non-classical path since
we know the analytic expression for them, but we analyse here only
the Gouy phase effect which can be measured for matter waves.
Moreover it is possible to tune  the parameters $t$ and $\tau$,
$\sigma_{0}$, $\beta$ and $d$ in order to study a specific phase
contribution.

To construct the graphic of the intensity and the Sorkin parameter
$\kappa$ we consider an electron wave with the following parameters:
$m=9.11\times10^{-31}\;\mathrm{kg}$, $d=650\;\mathrm{nm}$,
$\beta=62\;\mathrm{nm}$, $\sigma_{0}=62\;\mathrm{nm}$,
$t=18\;\mathrm{ns}$, $\tau=15\;\mathrm{ns}$. Using these parameters
as input we find $\epsilon=0.492ns$. In Fig. 2(a) we show the
normalized intensity $I_{n}$ as a function of $x$ which shows the
shape of intensity at far field or Fraunhofer theory as similarly
observed in \cite{Sinha2,Raedt}. In Fig. 2(b) we show the Sorkin
parameter $\kappa$ as a function of $x$ in which we consider (solid
line) and we do not consider (dotted line) the Gouy phase effect. In
accordance with \cite{Sinha2} we find that the quantity $\kappa$ is
of order $10^{-8}$ which corroborates our simplified analysis.
Moreover we verify numerically that the factor
$|\psi_{nc}(x,t,\tau)|^{2}$ does not change the parameter $\kappa$
significantly, the main contributions coming from the crossed terms
which contain the Gouy phase.

\begin{figure}[!htb]
\centering
\includegraphics[width=4.0 cm]{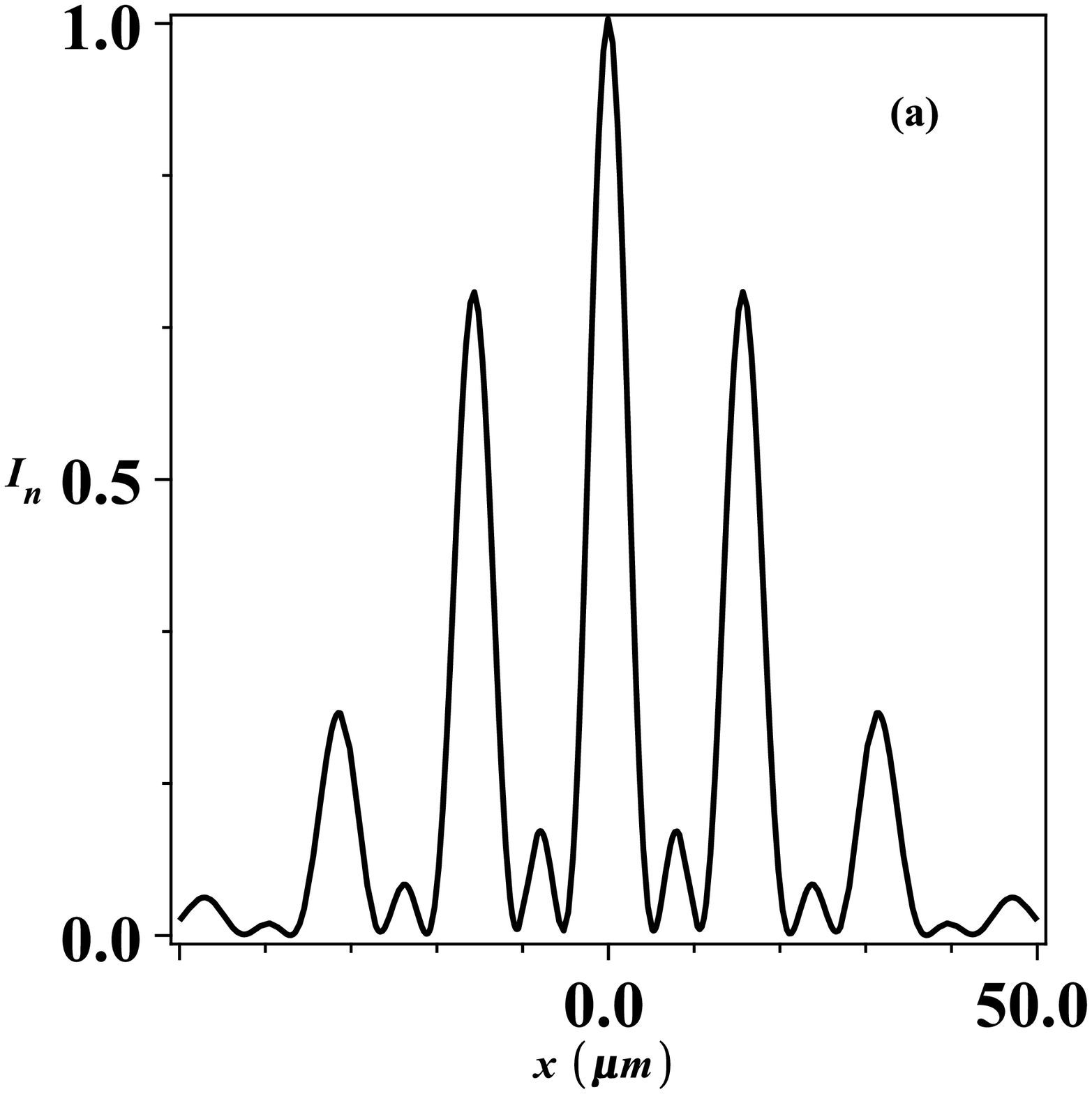}
\includegraphics[width=4.0 cm]{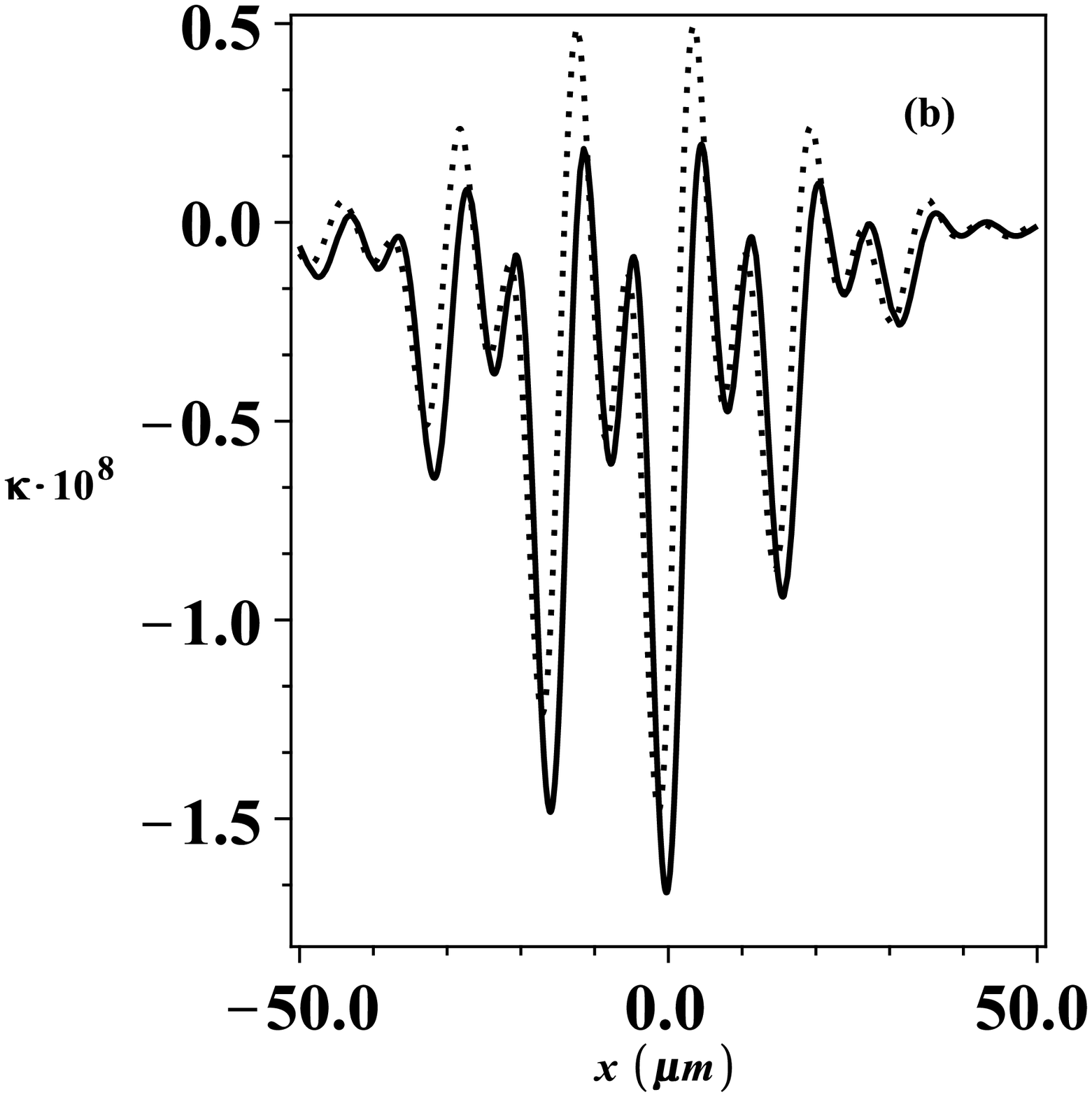}
\caption{(a) Normalized intensity as a function of $x$. (b) Sorkin
parameter $\kappa$ as a function of $x$. For solid line we consider
and for dotted line we do not consider the Gouy phase difference.
}\label{int_rel}
\end{figure}

In Fig. 3 we show the behavior of $|\kappa|$ as function of $x$ and
$\tau$. We observe that it has a maximum for a given value of $x$
and $\tau$. The existence of a maximum enable us choose a set of
value of parameters that produce a value of $\kappa$ which can be
more accessible to be measured. The existence of a maximum value for
this parameter as a function of the quantities involved in the
experimental apparatus was previously observed in \cite{Sinha3}. As
we can see this maximum occurs around $x=0$. Next we will explore
the Gouy phase effect to estimate the parameter $\kappa$ for $x=0$
since for this position some phases disappear making the Gouy phase
effect most evident.

\begin{figure}[!h]
\centering
\includegraphics[width=8.0 cm]{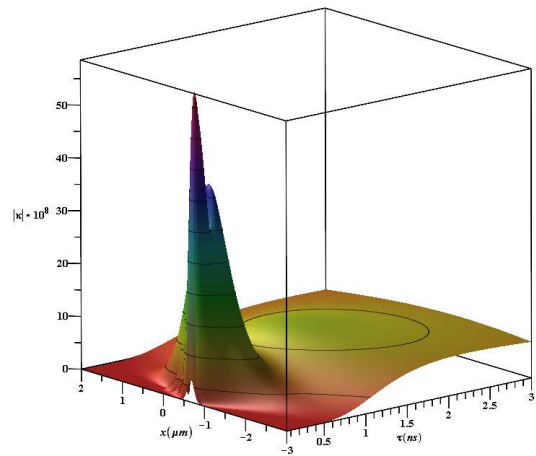}
\caption{Absolute value of Sorkin parameter $\kappa$ as a function
of $\tau$ and $x$ for $t$ fixed. It presents a maximum value for a
given value of $x$ and $\tau$.}\label{factorK1}
\end{figure}

In figure 4(a) we show the Gouy phase of classical path (red line)
and non-classical path (black line) as a function of $\tau$ for the
same parameters used above. We can observe that the absolute value
of the Gouy phase for the classical path decreases whereas for the
non-classical path it increases as the time propagation $\tau$
grows. This change affects the parameter $\kappa$. To observe such
effect, in Fig. 4(b) we show the absolute value of the parameter
$\kappa$ as a function of $\tau$ for $x=0$.

\begin{figure}[htp]
\centering
\includegraphics[width=4.0 cm]{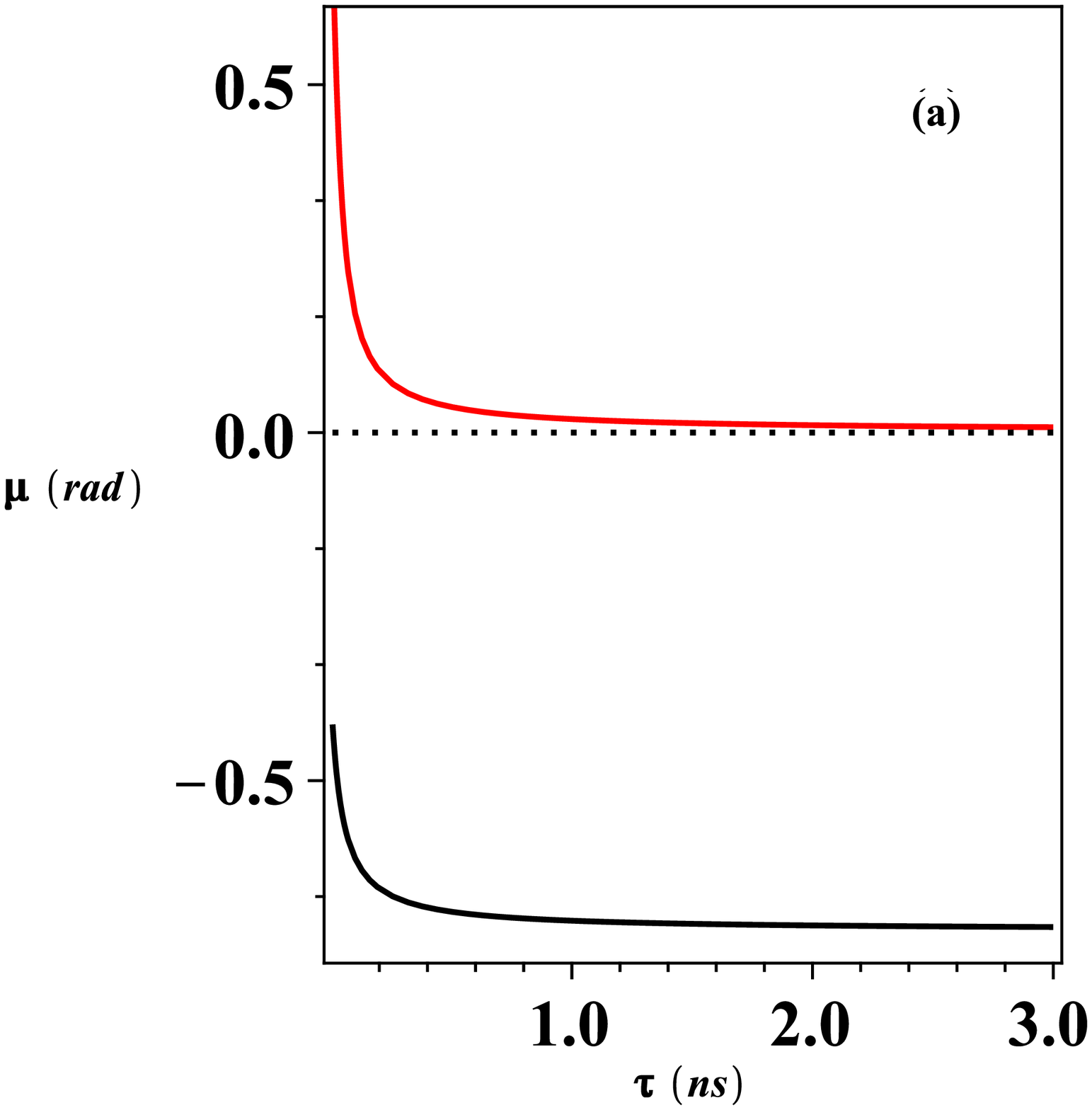}
\includegraphics[width=4.0 cm]{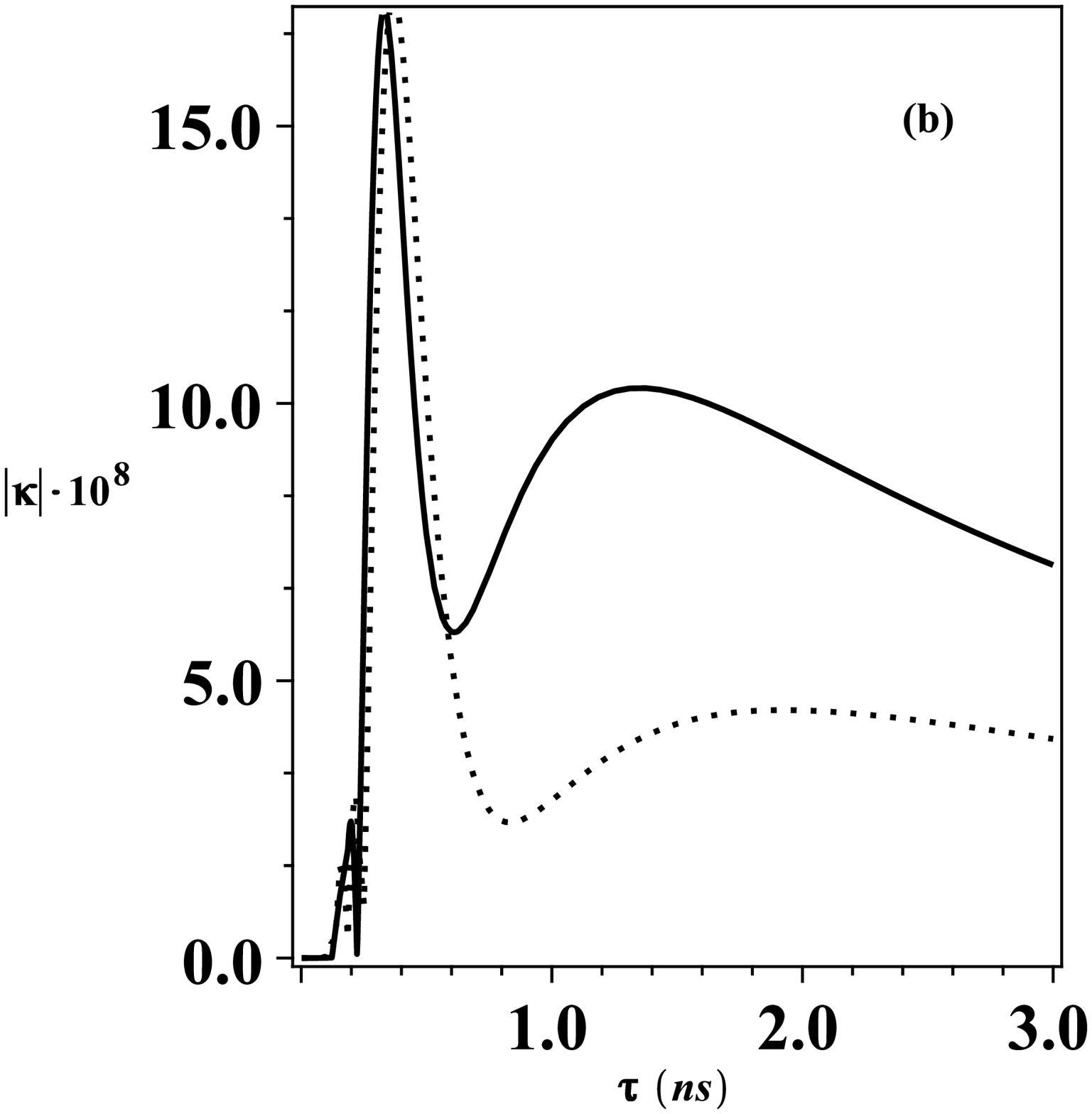}
\caption{(a) Gouy phase difference as a function of $\tau$ and $t$
fixed. (b) Absolute value of Sorkin parameter $\kappa$ as a function
of $\tau$ at $x=0$ and $t$ fixed.}\label{factorK1}
\end{figure}

The behavior of the parameter $\kappa$ as a function of $\tau$ is
similar with that obtained as a function of the distance of the
triple-slit to the screen in \cite{Sinha3}. For a given value of
$\tau$ it has a peak as in Ref. \cite{Sinha3}. It is noteworthy that
an exact solution for $\kappa$ depends on the Gouy phase. In order
to evaluate the effect of the Gouy phase on the absolute value of
the parameter $\kappa$, we calculate for point $x=0$ the percentage
error which is defined by
$(||\kappa|-|\kappa^{\prime}||/|\kappa|)\times100\%$, where for
$|\kappa|$ we consider the Gouy phase difference which corresponds
to the exact value and for $|\kappa^{\prime}|$ we neglected the Gouy
phase difference. Choosing $\tau=2\;\mathrm{ns}$  the percentage
error is $51.5\%$. Therefore if the Gouy phase is neglected the
parameter $\kappa$ is misestimated. As we can observe in figure 4(a)
for $\tau=2\;\mathrm{ns}$ the Gouy phase of classical path tends to
zero and the Gouy phase difference is due to the phase of the
non-classical path contribution , i.e.,
$\mu_{c}-\mu_{nc}\approx|\mu_{nc}|\approx0.719\;\mathrm{rad}$. If
one uses these parameters one measures the Gouy phase difference as
a signature of non-classical path contributions.

\section{IV. Concluding remarks}

We studied the effects of non-classical path in the interference
pattern in a triple-slit experiment. We solved exactly a one
dimensional model of propagation through a triple-slit and found
analytical expressions for the wavefunctions of classical and
non-classical paths. We obtained an exactly solution for the Sorkin
parameter $\kappa$ used to estimate the effect of non-classical
path. The value of $\kappa$ for electron waves is consistent with
other results previously obtained for it which make our model
reasonably good to study the existence of non-classical path. We
used the uncertainty in momentum to estimate the inter slits
propagation linking the existence of non-classical path with the
uncertainty principle which is as intuitive as to appeal to
Feynman's path integral formalism. The Gouy phase of classical and
non-classical paths are different which contribute significantly for
the value of $\kappa$. We observed the changing in the behavior of
$\kappa$ as a consequence of the Gouy phase difference for electron
waves. We estimated the percentage error in the absolute value of
parameter $\kappa$ as a consequence of the Gouy phase difference for
$x=0$ and $\tau=2\;\mathrm{ns}$ and found $51.5\%$. We conclude,
from the enormous discrepancy found, that the Gouy phase difference
can not be neglected in the three-slit interference if non-classical
paths are presents. We expected that our results which connect the
Sorkin parameter and Gouy phase must be further useful to detect
non-classical path's effect by measuring the Gouy phase. \vskip1.0cm
\begin{acknowledgments}
The authors would like to thank CNPq-Brazil for financial support.
I. G. da Paz thanks support from the program PROPESQ (UFPI/PI) under
grant number PROPESQ 23111.011083/2012-27.
\end{acknowledgments}

 \section{appendix 1: Formulae for interference parameters}

 In the following we display the complete expressions of terms in eqs. (\ref{psi1}), (\ref{psi2}), (\ref{psi3}) and (\ref{psi_nc}):

\bq A&=&\frac{m}{2\hbar\sqrt{\sqrt{\pi}t\tau
\sigma_{0}}}\Bigg[\left(\frac{m^{2}}{4\hbar^{2}t\tau}-\frac{1}{4\beta^{2}\sigma_{0}^{2}}\right)^{2}
\nonumber \\&+&
\frac{m^{2}}{16\hbar^{2}}\left(\frac{1}{\beta^{2}t}+\frac{1}{\sigma_{0}^{2}t}+\frac{1}{\sigma_{0}^{2}\tau}\right)^{2}\Bigg]^{-\frac{1}{4}},
\eq

\begin{equation}
C_{1}=\frac{\frac{m^{2}}{\hbar^{2}\tau^{2}}
{\cal{A}}}{4\left[{\cal{A}}^{2}+{\cal{B}}^{2}\right]},
\end{equation}
\begin{equation}
C_{2}=\frac{\frac{2md}{\hbar\tau\beta^{2}}{\cal{B}}}{4\left[{\cal{A}}^2+{\cal{B}}^2\right]},
\end{equation}
$$
{\cal{A}}=\left(\frac{1}{2\beta^{2}}+
\frac{m^{2}\sigma_{0}^{2}}{2(\hbar^{2}t^{2}+m^{2}\sigma_{0}^{4})}\right)
$$
$$
{\cal{B}}=\left(\frac{m^{3}\sigma_{0}^{4}}{2\hbar
t(\hbar^{2}t^{2}+m^{2}\sigma_{0}^{4})}-\frac{m}{2\hbar
t}-\frac{m}{2\hbar \tau}\right)
$$

\begin{equation}
C_{3}=-\frac{d^{2}}{2\beta^{2}}+\frac{\hbar^{2}\tau^{2}d^{2}}{m^{2}\beta^{2}}C_{1},\;\;\gamma=\frac{2d\hbar\tau}{m\beta^{2}}C_{1},\nonumber
\end{equation}
\begin{equation}
\alpha=\frac{mx^{2}}{2\hbar\tau}+\frac{m\beta^{2}}{2\hbar\tau}C_{2},\;\;\theta_{C}=\frac{\hbar\tau
d}{2m\beta^{2}}C_{2},
\end{equation}

\begin{equation}
\mu_{c}(t,\tau)=-\frac{1}{2}\arctan\left[\frac{t+\tau(1+\frac{\sigma_{0}^{2}}{\beta^{2}})}{\tau_{0}(1-\frac{t\tau\sigma_{0}^{2}}{\tau_{0}^{2}\beta^{2}})}\right],
\end{equation}

\begin{equation}
A_{nc}=\sqrt{\frac{m^{3}\sqrt{\pi}}{16\hbar^{3}\tau
t\epsilon\sigma_{0}\sqrt{z_{R}^{2}+z_{I}^{2}}}},
\end{equation}

\begin{equation}
C_{1nc}=\frac{m^{2}z_{3R}}{4\hbar^{2}\tau^{2}(z_{3R}^{2}+z_{3I}^{2})},
\end{equation}

\begin{equation}
C_{2nc}=\frac{m^{3}dz_{6I}}{32\hbar^{3}\beta^{2}\tau\epsilon^{2}(z_{6R}^{2}+z_{6I}^{2})}+\frac{mdz_{3I}}{2\hbar\tau\beta^{2}(z_{3R}^{2}+z_{3I}^{2})},
\end{equation}

\begin{eqnarray}
C_{3nc}&=&\frac{d^{2}z_{1R}}{4\beta^{4}(z_{1R}^{2}+z_{1I}^{2})}-\frac{m^{2}d^{2}z_{4R}}{64\beta^{4}\hbar^{2}\epsilon^{2}(z_{4R}^{2}+z_{4I}^{2})}\nonumber \\
&+&\frac{m^{4}d^{2}z_{5R}}{4^{5}\hbar^{4}\beta^{4}\epsilon^{4}(z_{5R}^{2}+z_{5I}^{2})}
+\frac{m^{2}d^{2}z_{6R}}{32\beta^{4}\epsilon^{2}\hbar^{2}(z_{6R}^{2}+z_{6I}^{2})}
\nonumber
\\&+&\frac{d^{2}z_{3R}}{4\beta^{4}(z_{3R}^{2}+z_{3I}^{2})}-\frac{d^{2}}{\beta^{2}},
\end{eqnarray}

\begin{equation}
\alpha_{nc}=\frac{mx^{2}}{2\hbar\tau}+\frac{m^{2}z_{3I}}{4\hbar^{2}\tau^{2}(z_{3R}^{2}+z_{3I}^{2})},
\end{equation}

\begin{equation}
\gamma_{nc}=\frac{m^{3}dz_{6R}}{32\hbar^{3}\beta^{2}\tau\epsilon^{2}(z_{6R}^{2}+z_{6I}^{2})}+\frac{mdz_{3R}}{2\hbar\tau\beta^{2}(z_{3R}^{2}+z_{3I}^{2})},
\end{equation}

\begin{eqnarray}
\theta_{nc}&=&\frac{d^{2}z_{1I}}{4\beta^{4}(z_{1R}^{2}+z_{1I}^{2})}-\frac{m^{2}d^{2}z_{4I}}{64\beta^{4}\hbar^{2}\epsilon^{2}(z_{4R}^{2}+z_{4I}^{2})}\nonumber
\\&+&
\frac{m^{4}d^{2}z_{5I}}{4^{5}\hbar^{4}\beta^{4}\epsilon^{4}(z_{5R}^{2}+z_{5I}^{2})}
+\frac{m^{2}d^{2}z_{6I}}{32\beta^{4}\epsilon^{2}\hbar^{2}(z_{6R}^{2}+z_{6I}^{2})}\nonumber
\\&+& \frac{d^{2}z_{3I}}{4\beta^{4}(z_{3R}^{2}+z_{3I}^{2})},
\end{eqnarray}

\section{appendix 2: Gouy phase components}


In the following we present the full expression of the terms in
equation (\ref{ncgouy}).

\begin{eqnarray}
z_{R}&=&(z_{0R}z_{1R}-z_{0I}z_{1I})(z_{2R}z_{3I}+z_{2I}z_{3R})+\nonumber
\\&+&(z_{0R}z_{1I}+z_{0I}z_{1R})(z_{2R}z_{3R}-z_{2I}z_{3I}),
\end{eqnarray}

\bq
z_{I}&=&(z_{0R}z_{1R}-z_{0I}z_{1I})(z_{2R}z_{3R}-z_{2I}z_{3I})\nonumber\\&-&(z_{0R}z_{1I}+z_{0I}z_{1R})(z_{2R}z_{3I}+z_{2I}z_{3R}),
\eq

\begin{equation}
z_{0R}=\frac{1}{2\sigma_{0}^{2}},\;\;z_{0I}=-\frac{m}{2\hbar t},
\end{equation}
\begin{equation}
z_{1R}=\frac{1}{2\beta^{2}}+\frac{m^{2}z_{0R}}{4\hbar^{2}
t^{2}(z_{0R}^{2}+z_{0I}^{2})},\;\;
\end{equation}
\begin{equation}
z_{1I}=-\left(\frac{m}{4\hbar \epsilon}+\frac{m}{2\hbar
t}+\frac{m^{2}z_{0I}}{4\hbar^{2}t^{2}(z_{0R}^{2}+z_{0I}^{2})}\right),
\end{equation}
\begin{equation}
z_{2R}=\frac{1}{2\beta^{2}}+\frac{m^{2}z_{1R}}{16\hbar^{2}\epsilon^{2}(z_{1R}^{2}+z_{1I}^{2})},
\end{equation}
\begin{equation}
z_{2I}=-\left(\frac{m}{2\hbar\epsilon}+\frac{m^{2}z_{1I}}{16\hbar^{2}\epsilon^{2}(z_{1R}^{2}+z_{1I}^{2})}\right),
\end{equation}

\begin{equation}
z_{3R}=\frac{1}{2\beta^{2}}+\frac{m^{2}z_{2R}}{16\hbar^{2}\epsilon^{2}(z_{2R}^{2}+z_{2I}^{2})},
\end{equation}

\begin{equation}
z_{3I}=-\left(\frac{m}{2\hbar
\tau}+\frac{m}{4\hbar\epsilon}+\frac{m^{2}z_{2I}}{16\hbar^{2}\epsilon^{2}(z_{2R}^{2}+z_{2I}^{2})}\right),
\end{equation}

\begin{equation}
z_{4R}=z_{1R}^{2}z_{2R}-z_{1I}^{2}z_{2R}-2z_{1R}z_{1I}z_{2I},
\end{equation}

\begin{equation}
z_{4I}=z_{1R}^{2}z_{2I}-z_{1I}^{2}z_{2I}+2z_{1R}z_{1I}z_{2R},
\end{equation}

\begin{eqnarray}
&&z_{5R}=z_{3R}\big(z_{1R}^{2}z_{2R}^{2}-z_{1R}^{2}z_{2I}^{2}-z_{1I}^{2}z_{2R}^{2}+z_{1I}^{2}z_{2I}^{2}\nonumber
\\ &&-4z_{1R}z_{1I}z_{2R}z_{2I}\big)
-2z_{3I}\big(z_{1R}^{2}z_{2R}z_{2I}-z_{1I}^{2}z_{2R}z_{2I}\nonumber
\\&&+z_{1R}z_{1I}z_{2R}^{2}-z_{1R}z_{1I}z_{2I}^{2}\big),
\end{eqnarray}
\begin{eqnarray}
&&z_{5I}=z_{3I}(z_{1R}^{2}z_{2R}^{2}-z_{1R}^{2}z_{2I}^{2}-z_{1I}^{2}z_{2R}^{2}+z_{1I}^{2}z_{2I}^{2}
\nonumber \\
&&-4z_{1R}z_{1I}z_{2R}z_{2I})+2z_{3R}(z_{1R}^{2}z_{2R}z_{2I}\nonumber
\\&&-z_{1I}^{2}z_{2R}z_{2I}+z_{1R}z_{1I}z_{2R}^{2}-z_{1R}z_{1I}z_{2I}^{2}),
\end{eqnarray}
\begin{equation}
z_{6R}=z_{1R}z_{2R}z_{3R}-z_{1R}z_{2I}z_{3I}-z_{1I}z_{2R}z_{3I}-z_{1I}z_{2I}z_{3R},
\end{equation}

and

\begin{equation}
z_{6I}=z_{1R}z_{2R}z_{3I}+z_{1R}z_{2I}z_{3R}+z_{1I}z_{2R}z_{3R}-z_{1I}z_{2I}z_{3I}.
\end{equation}

\end{document}